\documentclass[twocolumn,showpacs,preprintnumbers,amsmath,amssymb,superscriptaddress]{revtex4}

\usepackage{graphicx}
\begin{document}
\bibliographystyle{prsty}
\title{Temperature-dependent photoemission spectral weight transfer and chemical potential shift in Pr$_{1-x}$Ca$_x$MnO$_3$ : Implications for charge density modulation}

\author{K. Ebata}
\affiliation{Department of Physics and Department of Complexity Science and Engineering, University of Tokyo, 7-3-1 Hongo, Bunkyo-ku, Tokyo 113-0033, Japan}
\author{M. Hashimoto}
\affiliation{Department of Physics and Department of Complexity Science and Engineering, University of Tokyo, 7-3-1 Hongo, Bunkyo-ku, Tokyo 113-0033, Japan}
\author{K. Tanaka}
\affiliation{Department of Physics and Department of Complexity Science and Engineering, University of Tokyo, 7-3-1 Hongo, Bunkyo-ku, Tokyo 113-0033, Japan}
\author{A. Fujimori}
\affiliation{Department of Physics and Department of Complexity Science and Engineering, University of Tokyo, 7-3-1 Hongo, Bunkyo-ku, Tokyo 113-0033, Japan}
\author{Y. Tomioka}
\affiliation{Correlated Electron Research Center (CERC), National Institute of Advanced Industrial Science and Technology (AIST), Tsukuba 305-8562, Japan}
\author{Y. Tokura}
\affiliation{Correlated Electron Research Center (CERC), National Institute of Advanced Industrial Science and Technology (AIST), Tsukuba 305-8562, Japan}
\affiliation{Department of Applied Physics, University of Tokyo, Bunkyo-ku, Tokyo 113-8656, Japan}
\affiliation{Spin Superstructure Project, Exploratory Research for Advanced Technology (ERATO), Japan Science and Technology Corporation (JST), Tsukuba 305-8562, Japan}
\date{\today}

\begin{abstract}
We have studied the temperature dependence of the photoemission spectra of Pr$_{1-x}$Ca$_x$MnO$_3$ (PCMO) with $x=0.25$, 0.3 and 0.5. For $x=0.3$ and 0.5, we observed a gap in the low-temperature CE-type charge-ordered (CO) phase and a pseudogap with a finite intensity at the Fermi level ($E_F$) in the high-temperature paramagnetic insulating (PI) phase. Within the CO phase, the spectral intensity near $E_F$ gradually increased with temperature. These observations are consistent with the results of Monte Carlo simulations on a model including charge ordering and ferromagnetic fluctuations [H. Aliaga {\it et al.} Phys. Rev. B {\bf 68}, 104405 (2003)]. For $x=0.25$, on the other hand, little temperature dependence was observed within the low-temperature ferromagnetic insulating (FI) phase and the intensity at $E_F$ remained low in the high-temperature PI phase. We attribute the difference in the temperature dependence near $E_F$ between the CO and FI phases to the different correlation lengths of orbital order between both phases. Furthermore, we observed a chemical potential shift with temperature due to the opening of the gap in the FI and CO phases. The doping dependent chemical potential shift was recovered at low temperatures, corresponding to the disappearance of the doping dependent change of the modulation wave vector. Spectral weight transfer with hole concentration was clearly observed at high temperatures but was suppressed at low temperatures. We attribute this observation to the fixed periodicity with hole doping in PCMO at low temperatures.
\end{abstract}


\maketitle
\section{Introduction}
Recently, manganites with the perovskite structure have attracted considerable attention due to the discovery of the colossal magnetoresistance (CMR), and the complicated interplay between the spin, charge, and orbital degrees of freedom \cite{Tokura5}. Many experimental and theoretical studies have indicated that the inhomogeneous nature arising from competition between the ferromagnetic (FM) metallic and antiferromagnetic (AFM) insulating states leads to a large change in the resistivity as a function of temperature, magnetic field, chemical pressure and so on.
\begin{figure}
\begin{center}
\includegraphics[width=6.5cm]{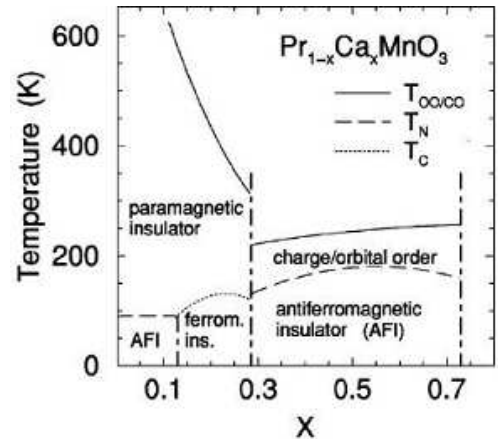}
\caption{Electronic phase diagram of Pr$_{1-x}$Ca$_x$MnO$_3$ \cite{Zimmermann2, Tomioka, Jirak}.}
\label{phasePCMO1}
\end{center}
\end{figure}
The presence of coexsisting clusters of FM charge-disordered and AFM charge-ordered (CO) states in (La$_{1-y}$Pr$_y$)$_{1-x}$Ca$_x$MnO$_3$ was clearly shown by means of electron microscopy \cite{Uehara}. Fluctuations between the FM and CO phases were found in the paramagnetic insulating (PI) phase of Pr$_{1-x}$Ca$_x$MnO$_3$ (PCMO) by means of neutron scattering and x-ray resonant scattering studies \cite{Kajimoto, Zimmermann}. From the studies of photoemission spectroscopy (PES), pseudogap features with a finite intensity at the Fermi level ($E_F$) were observed in the PI phase of PCMO, consistent with the fluctuations involving both phases \cite{Ebata}. Monte Carlo simulations on a two-orbital model showed a temperature evolution of the density of states (DOS) from a clear gap due to the charge ordering at low temperatures to a pseudogap in the competing phases at high temperatures \cite{Aliaga}. The calculation also exhibited the gaps in the CO phase of the manganites were much larger than $k_B$$T_{CO}$ ($T_{CO}$ : CO transition temperature) \cite{Yunoki2}. Okimoto {\it et al.} \cite{Okimoto} showed the existence of an optical gap in the CO phase of PCMO. From valence-band PES and O $K$ edge x-ray absorption studies, Dalai {\it et al.} \cite{Dalai} revealed that the charge-transfer energy of PCMO takes a large value, consistent with the strong charge localization in PCMO. Furthermore, the gap in the CO phase was detected in Nd$_{1-x}$Sr$_x$MnO$_3$ (NSMO) by means of PES and tunneling spectroscopy \cite{Sekiyama2, Biswas}. Small bandwidth systems such as PCMO exhibit a so-called CE-type AFM CO state and CMR effect becomes remarkable owing to the collapse of the CO state under a magnetic field \cite{Tomioka}. PCMO has a particularly stable CO state over a wide hole concentration range as shown in the electronic phase diagram in Fig. 1 \cite{Zimmermann2, Tomioka, Jirak}. In addition, PCMO shows an insulator-to-metal transition under various external perturbations such as electric field \cite{Asamitsu}, high pressure \cite{Moritomo}, and light irradiation \cite{Miyano, Fiebig}.

In this paper, we report on the results of a detailed temperature-dependent PES study of PCMO, where CMR is manifested for $x \geq 0.3$. We observed a pseudogap behavior with a finite DOS at $E_F$ in the high-temperature PI phase and a clear gap opening in the low-temperature CO phase, indicating competition between the FM and CO phases. Also, the spectra in the CO phase showed a remarkable temperature dependence. For $x<0.3$, on the other hand, the DOS at $E_F$ was found to be low in the PI phase and the spectra in the ferromagnetic insulating (FI) phase showed no significant temperature dependence. The gap was opened in the FI and CO phases and an upward chemical potential shift with decreasing temperature was observed. When the charge modulation wave vector as a function of hole concentration in the CO phase of PCMO became constant at low temperatures, the doping dependent chemical potential shift became large and monotonous, probably related to the spectral weight transfer near the $E_F$ with hole doping.

\section{Experimental}
Single crystals of PCMO with the Ca concentrations of $x=0.25$, 0.3, and 0.5 were grown by the floating-zone method. The growth techniques and transport properties of the crystals were described in Ref.\cite{Tomioka}. Ultraviolet photoemission spectroscopy (UPS) measurements were performed using the photon energy of $h\nu =$ 21.2 eV (He I). All the photoemission measurements were performed under the base pressure of $\sim 10^{-10}$ Torr at 80-300 K. The samples were repeatedly scraped {\it in situ} at several temperatures with a diamond file to obtain clean surfaces. The scraping was made until a bump around 9-10 eV, which was attributed to surface contamination, decreased and the valence-band spectra did not change with further scraping. Photoelectrons were collected using a Scienta SES-100 electron-energy analyzer. The energy resolution was about 15 meV. 

\section{Results and discussion}
\begin{figure}
\begin{center}
\includegraphics[width=7cm]{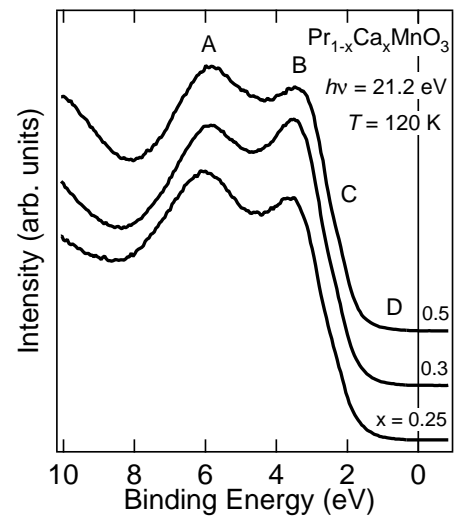}
\caption{Valence-band photoemission spectra of Pr$_{1-x}$Ca$_x$MnO$_3$.}
\label{valence1}
\end{center}
\end{figure}
\begin{figure}
\begin{center}
\includegraphics[width=9.2cm]{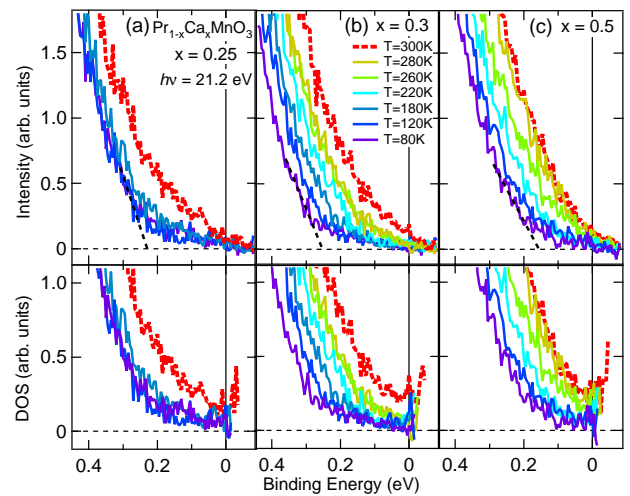}
\caption{(Color online) Temperature-dependent photoemission spectra near $E_F$ of Pr$_{1-x}$Ca$_x$MnO$_3$, (a) $x=0.25$; (b) $x=0.3$; (c) $x=0.5$. The bottom panels show the density of states (DOS) obtained by dividing the spectrum in the upper panels by the Fermi Dirac function.}
\label{valence2}
\end{center}
\end{figure}
Figure 2 shows PES spectra in the valence-band region of PCMO at 120 K. The spectra consisted of four main structures labeled A, B, C, and D, which are assigned to the Mn 3$d$-O 2$p$ bonding state, the non-bonding O 2$p$ state, the Mn 3$d$ $t_{2g}$ plus the Pr 4$f$ states, and the Mn 3$d$ $e_g$ state, respectively \cite{Ebata}.

Figure 3(a)-(c) show the temperature-dependent spectra near $E_F$ for the three compositions. The spectra have been normalized to the integrated intensity in the energy range from binding energy $E_B$ $=1.2$ eV to above $E_F$.
For $x=0.3$ and 0.5, below $T_{CO}$ we observed the opening of a gap due to the CO phase. The magnitude of the CO gap was estimated to be about 250 meV for $x=0.3$ and 150 meV for $x=0.5$ at 80 K, consistent with the optical gap energy estimated to be about 180 meV at 10 K for PCMO with $x=0.4$ \cite{Okimoto}. By PES measurements, Sekiyama {\it et al.} \cite{Sekiyama2} found a CO gap in NSMO ($x=0.5$), too, and estimated the magnitude to be about 100 meV. The magnitude of the CO gaps in manganites is much larger than $k_B$$T_{CO}$, in agreement with the prediction of the Monte Carlo simulation on the two-orbital model \cite{Yunoki2}. On the other hand, for $x=0.25$, there was a gap structure at all temperatures and the FI gap was estimated to be about 230 meV \cite{shift}.

We shall discuss the temperature dependence of the spectra near $E_F$ of PCMO within the CO and FI phases in more detail. The spectra showed a remarkable temperature dependence within the CO phase for $x=0.3$ and 0.5 while little temperature dependence was observed within the FI phase for $x=0.25$. From an x-ray resonant scattering study, Zimmermann {\it et al.} \cite{ Zimmermann2} found that for PCMO with $x=0.4$ and 0.5 while long-range charge order was present, short-range orbital order was also realized. On the other hand, for PCMO with $x=0.25$, long-range orbital order was observed with no indication of charge ordering \cite{Zimmermann2}. As a result, the correlation lengths of orbital order had temperature dependence for $x=0.4$ and 0.5 but not for $x=0.25$ \cite{Zimmermann2}. Therefore, the difference in the temperature evolution of the gap structures between the FI and CO phases may be related to the different temperature dependence of the correlation lengths of orbital order in both phases. Recently, the CO gap in NSMO ($x=0.5$) was also found to show temperature dependence by means of PES and tunneling spectroscopy \cite{Sekiyama2, Biswas}.

In order to eliminate the effect of the temperature dependence of the Fermi Dirac function and to extract the intrinsic temperature-dependent changes of the DOS, we have divided each PES spectrum by the Fermi Dirac function at each temperature as shown in the bottom panels of Fig. 3. While the insulating gaps were opened at all temperatures for $x=0.25$, finite intensity at $E_F$ was observed in the PI phase of $x=0.3$ and 0.5 \cite{Ebata}. The presence of the pseudogap, i.e., a gap-like feature with a finite DOS at $E_F$, at high temperatures and the real gap opening in the CO phase for $x=0.3$ and 0.5 are in good qualitative agreement with Monte Carlo calculation with competition between the CO and FM phases \cite{Aliaga}. It was also shown by the neutron scattering and x-ray resonant scattering experiments that both FM and CO fluctuations were present above $T_{CO}$ in PCMO \cite{Kajimoto, Zimmermann}. We therefore consider that the pseudogaps appear when the insulating and metallic phases are competing with each other.

\begin{figure}
\begin{center}
\includegraphics[width=6.5cm]{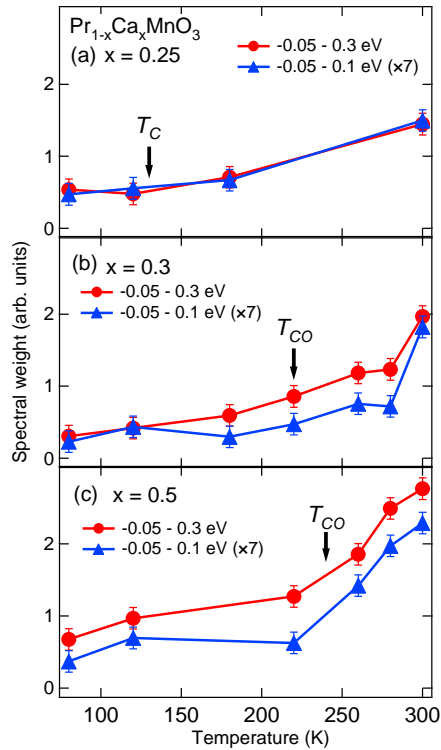}
\caption{(Color online) Spectral weight integrated from $E_B$ $=-0.05$ eV to 0.3 eV and that from $E_B$ $=-0.05$ eV to 0.1 eV in Pr$_{1-x}$Ca$_x$MnO$_3$ as functions of temperature, (a) $x=0.25$; (b) $x=0.3$; (c) $x=0.5$. $T_C$ and $T_{CO}$ denote the Curie temperature and the charge ordering temperature, respectively. All the spectra have been normalized to the integrated intensity in the wide energy range of $E_B$ $=1.2$ eV to $\sim$ above $E_F$.}
\label{weight}
\end{center}
\end{figure}
In Fig. 4, we have plotted the spectral weight integrated from $E_B$ $=-0.05$ eV to 0.3 eV and that from $E_B$ $=-0.05$ eV to 0.1 eV in PCMO as a function of temperature. The gradual increase of spectral weight near $E_F$ with temperature rather than a sudden jump at $T_{CO}$ is probably related with the thermally activated hopping of $e_g$ electrons at high temperatures as reported for Bi$_{1-x}$Ca$_x$MnO$_3$ \cite{Bao, Liu}.
In the previous PES studies of the FM phase of La$_{1-x}$Sr$_x$MnO$_3$ (LSMO) ($x$ = 0.4), the decreasing spectral weight at $E_F$ with increasing temperature was observed and was attributed to spectral weight transfer from the coherent to the incoherent parts with increasing temperature \cite{Sarma2, Saitoh2}. The temperature-dependent change of the spectral weight at $E_F$ in PCMO which exhibits CO phase was opposite to that in LSMO ($x$ = 0.4). The different temperature evolution of the DOS for the different types of ground states may be related to the two different types of CMR: (1) CMR which occurs in the high temperatures paramaganetic regime as in the case of LSMO, where the system undergoes a PI to FM transition upon cooling; (2) CMR which occurs in the CO phase at low temperature due to the melting of the CO state under magnetic field as in the case of PCMO.

\begin{figure}
\begin{center}
\includegraphics[width=8.4cm]{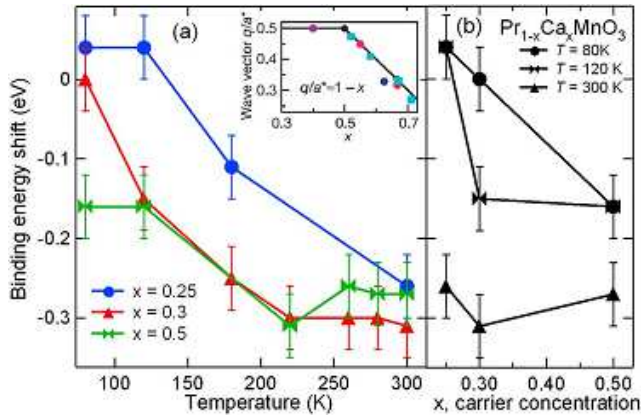}
\caption{(Color online) Binding energy shift as a function of temperature (a) and carrier concentration (b) in Pr$_{1-x}$Ca$_x$MnO$_3$. Inset shows the wave vector of the modulation versus carrier concentration $x$ for Pr$_{1-x}$Ca$_x$MnO$_3$ and La$_{1-x}$Ca$_x$MnO$_3$ at low temperatures \cite{Milward}.}
\label{shift}
\end{center}
\end{figure}
\begin{figure}
\begin{center}
\includegraphics[width=6.5cm]{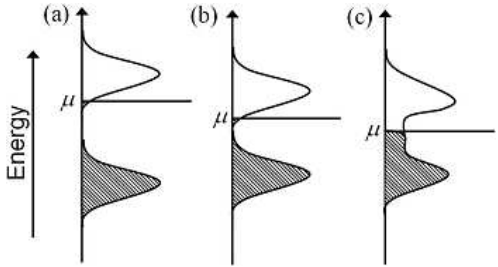}
\caption{Schematic pictures of the temperature-dependent DOS and chemical potential shift in the charge-ordered (CO) composition range of Pr$_{1-x}$Ca$_x$MnO$_3$. (a) Low-temperature CO phase; (b) intermediate-temperature CO phase; (c) high-temperature paramagnetic insulating (PI) phase. $\mu$ denotes chemical potential.}
\label{DOS_T}
\end{center}
\end{figure}
Next, we deduce the composition and temperature dependent chemical potential shift and discuss its implications of the charge modulations and fluctuations in PCMO.
In order to deduce the amount of the chemical potential shift, we have used the peak position of the Mn 3$d$-O 2$p$ bonding state because the peak position of the non-bonding O 2$p$ state was affected by the Pr 4$f$ state whose intensity changes with $x$. In Fig. 5, we have plotted the shift as a function of temperature and carrier concentration in PCMO. Panel (a) shows that an upward chemical potential shift occurred with decreasing temperature. We attribute it to the opening of gap in the FI and CO phases of PCMO. In Fig. 6, a schematic diagram for the temperature-dependent chemical potential shift in the CO composition range of PCMO is shown. In the CO phase, the CO gap is opened corresponding to the charge ordering in the CE-type AFM state. If the hole concentration is less than $x=0.5$, the upper band is partially occupied by electrons, and the chemical potential is located near its bottom. Therefore, the chemical potential is shifted upward with decreasing temperature. For $x \geq 0.3$, we also observed the suppression of the carrier concentration-dependent shift at high temperatures as shown in Fig. 5(b). We consider that the suppression of the shift is related to charge self-organization such as fluctuating stripe formation in the CO composition range as inferred from the chemical potential shift determined from the core-level photoemission spectra \cite{Ebata}. In fact, the structural modulation wave vector in PCMO changes with carrier concentration at high temperatures in the CO composition range \cite{Milward}. At low temperatures, on the other hand, a smooth chemical potential shift with carrier concentration was observed. This may be related with the fixed periodicity of the CO state in PCMO at low temperatures for $x < 0.5$, as shown in the inset of Fig. 5 \cite{Milward}.

\begin{figure}
\begin{center}
\includegraphics[width=9.2cm]{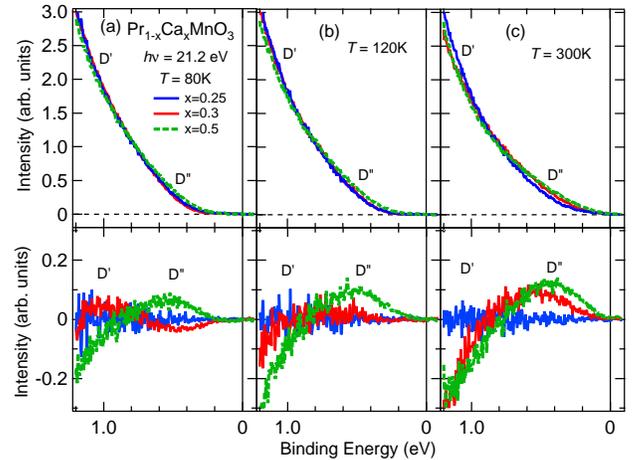}
\caption{(Color online) Hole-concentration dependent photoemission spectra near $E_F$ of Pr$_{1-x}$Ca$_x$MnO$_3$, (a) $T=80$ K; (b) $T=120$ K; (c) $T=300$ K. The bottom panels show the difference spectra obtained by subtracting the spectra of $x=0.25$.}
\label{valence3}
\end{center}
\end{figure}
In Fig. 7, we have plotted the valence-band spectra of the three samples near $E_F$ as a function of hole concentration. Spectral weight transfer from D' to D" with hole concentration was clearly observed at $T=300$ K as repeated in Ref.\cite{Ebata}, as can be seen from the difference spectra in the bottom panel. This effect can be attributed to the change of periodicity in the fluctuating stripes with hole concentration at high temperatures \cite{Milward}. With decreasing temperature, the spectral weight transfer was gradually suppressed, concomitant with the recovery of the chemical potential shift. It is likely that the different behavior of the spectral weight transfer with hole concentration reflects the different evolution of the periodicity of the charge modulation in PCMO.

\section{Conclusion}
In conclusion, we have studied the temperature dependence of the DOS around $E_F$ in PCMO by means of PES and observed a clear gap opening in the CO phase and a pseudogap with a finite intensity at $E_F$ in the high-temperature PI phase for $x=0.3$ and 0.5. The spectra within the CO phase also showed significant temperature dependence, probably related to the temperature dependence of the short-range orbital ordering. The temperature-dependent redistribution of spectral weight for $x=0.3$ and 0.5 below and above $T_{CO}$ was in good qualitative agreement with the results of the Monte Carlo simulation on a model for manganite with competing CO and FM fluctuations. For $x=0.25$, a clear gap feature was found at all temperatures and the spectra within the FI phase showed little temperature dependence, which may be connected with the stable long-range orbital ordering. The temperature-dependent chemical potential shift was also observed in PCMO. We consider the shift to be related with the opening of the gap in the FI and CO phases. The hole-concentration dependent chemical potential shift was found to be suppressed at high temperatures, which we attribute to the charge self-organization such as stripes. Also, the spectral weight transfer with hole concentration was weakened with decreasing temperature, probably related to the recovery of the hole-concentration dependent chemical potential shift in PCMO at low temperatures.

\section{Acknowledgment}
This work was supported by a Grant-in-Aid for Scientific Research in Priority Area ``Invention of Anomalous Quantum Materials" from the Ministry of Education, Culture, Sports, Science and Technology, Japan.


\begin{thebibliography}{10}

\bibitem{Tokura5}
Y. Tokura, Rep. Prog. Phys {\bf 69},  797  (2006).

\bibitem{Zimmermann2}
M. v.~Zimmermann, C.~S. Nelson, J.~P. Hill, D. Gibbs, M. Blume, D. Casa, B.
  Keimer, Y. Murakami, C.-C. Kao, C. Venkataraman, T. Gog, Y. Tomioka, and Y.
  Tokura, Phys. Rev. B {\bf 64},  195133  (2001).

\bibitem{Tomioka}
Y. Tomioka, A. Asamitsu, H. Kuwahara, Y. Moritomo, and Y. Tokura, Phys. Rev. B
  {\bf 53},  R1689  (1996).

\bibitem{Jirak}
Z. Jirak, S. Krupica, Z. Simsa, M. Dlouha, and S. Vratislav, J. Magn. Magn.
  Mater {\bf 53},  153  (1985).

\bibitem{Uehara}
M. Uehara, S. Mori, C.~H. Chen, and S.-W. Cheong, Nature {\bf 399},  560
  (1999).

\bibitem{Kajimoto}
R. Kajimoto, T. Kakeshita, Y. Oohara, H. Yoshizawa, Y. Tomioka, and Y. Tokura,
  Phys. Rev. B {\bf 58},  R11837  (2003).

\bibitem{Zimmermann}
M. v.~Zimmermann, J.~P. Hill, D. Gibbs, M. Blume, D. Casa, B. Keimer, Y.
  Murakami, Y. Tomioka, and Y. Tokura, Phys. Rev. Lett. {\bf 83},  4872
  (2003).

\bibitem{Ebata}
K. Ebata, H. Wadati, M. Takizawa, A. Fujimori, A. Chikamatsu, H. Kumigashira,
  M. Oshima, Y. Tomioka, and Y. Tokura, Phys. Rev. B {\bf 74},  064419  (2006).

\bibitem{Aliaga}
H. Aliaga, D. Magnoux, A. Moreo, D. Poilblanc, S. Yunoki, and E. Dagotto, Phys.
  Rev. B {\bf 68},  104405  (2003).

\bibitem{Yunoki2}
S. Yunoki, T. Hotta, and E. Dagotto, Phys. Rev. Lett. {\bf 84},  3714  (2000).

\bibitem{Okimoto}
Y. Okimoto, Y. Tomioka, Y. Onose, Y. Otsuka, and Y.Tokura, Phys. Rev. B {\bf
  57},  R9377  (1998).

\bibitem{Dalai}
M.~K. Dalai, P. Pal, B.~R. Sekhar, N.~L. Saini, R.~K. Singhal, K.~B. Garg, B.
  Doyle, S. Nannarone, C. Martin, and F. Studer, Phys. Rev. B {\bf 74},  165119
   (2006).

\bibitem{Sekiyama2}
A. Sekiyama, S. Suga, M. Fujikawa, S. Imada, T. Iwasaki, K. Matsuda, T.
  Matsushita, K.~V. Kaznacheyev, A. Fujimori, H. Kuwahara, and Y. Tokura, Phys.
  Rev. B {\bf 59},  15528  (1999).

\bibitem{Biswas}
A. Biswas, A.~K. Raychaudhuri, R. Mahendiran, A. Guha, R. Mahesh, and C.~N.~R.
  Rao, J. Phys.: Condens. Mat. {\bf 9},  L355  (1997).

\bibitem{Asamitsu}
A. Asamitsu, Y. Tomioka, H. Kuwahara, and Y. Tokura, Nature {\bf 388},  50
  (1997).

\bibitem{Moritomo}
Y. Moritomo, H. Kuwahara, Y. Tomioka, and Y. Tokura, Phys. Rev. B {\bf 55},
  7549  (1997).

\bibitem{Miyano}
K. Miyano, T. Tanaka, Y. Tomioka, and Y. Tokura, Phys. Rev. Lett. {\bf 78},
  4257  (1997).

\bibitem{Fiebig}
M. Fiebig, K. Miyano, Y. Tomioka, and Y. Tokura, Science {\bf 280},  1925
  (1998).

\bibitem{shift}
Since for $x=0.25$ the spectra near $E_F$ was nearly unchanged within the low-temperature FI phase, the charging effect, which causes an energy shift towards higher binding energies, can be excluded. Also, charging effects for $x=0.3$ and 0.5 were unlikely because the resistivities at 80 K of the $x=0.3$ and 0.5 samples were lower than that of $x=0.25$ [Y. Tomioka, A. Asamitsu, H. Kuwahara, Y. Moritomo, and Y. Tokura, Phys. Rev. B {\bf 53},  R1689  (1996); R. Kajimoto, H. Mochizuki, H. Yoshizawa, S. Okamoto, and S. Ishihara, Phys. Rev. B {\bf 69},  054433  (2004)].

\bibitem{Bao}
W. Bao, J.~D. Axe, C.~H. Chen, and S.-W. Cheong, Phys. Rev. Lett. {\bf 78},  543
  (1997).

\bibitem{Liu}
H.~L. Liu, S.~L. Cooper, and S.-W. Cheong, Phys. Rev. Lett. {\bf 81},  4684
  (1998).

\bibitem{Sarma2}
D.~D. Sarma, N. Shanthi, S.~R. Krishnakumar, T. Saitoh, T. Mizokawa, A.
  Sekiyama, K. Kobayashi, A. Fujimori, E. Weschke, R. Meier, G. Kaindl, Y.
  Takeda, and M. Takano, Phys. Rev. B {\bf 53},  6873  (1996).

\bibitem{Saitoh2}
T. Saitoh, A. Sekiyama, K. Kobayashi, T. Mizokawa, A. Fujimori, D.~D. Sarma, Y.
  Takeda, and M. Takano, Phys. Rev. B {\bf 56},  8836  (1997).

\bibitem{Milward}
G.~C. Milward, M.~J. Calderon, and P.~B. Littlewood, Nature {\bf 433},  607
  (2005).

\end{thebibliography}
\end{document}